\documentclass[a4paper]{article}
\usepackage[T1]{fontenc}
\usepackage[utf8]{inputenc}
\usepackage[italian,english]{babel}
\usepackage[autostyle,italian=guillemets]{csquotes}
\usepackage{multicol}
\usepackage{microtype}
\usepackage{guit}
\usepackage{booktabs}
\usepackage{graphicx}
\usepackage{mathptmx}
\usepackage{amsfonts}
\usepackage{amssymb}
\usepackage{amsmath}
\usepackage{ifpdf}
\usepackage{dcolumn}
\usepackage{textcomp}
\usepackage{tabularx,array}
\usepackage{ulem}
\usepackage{placeins}
\usepackage{epstopdf}
\usepackage{etex}
\date{}

\title{New CCD Photometry Study of RV UMa} 
\author{Tasselli, D. \\ TS Corporation Srl - Department of Astronomy and Astrophysics \\ Regione Salamia, 10010 Andrate TO - Italy \\ E-mail:diego.tasselli@tscorporation.org} 
\begin{document}
 
 \maketitle
 \begin{abstract}
 \normalsize All available CCD observation of RV UMa have been analyzed to obtain an accurate mathematical description of the ligh variation. We discuss in this paper a new study of variable star RV UMa, a short period RRab star, in orther to determine through the light curve and the physical parameters, the presence of \textit{``Blazhko effect''}. The Star were observed for a total of 839 sessions shooting, and exhibits light curve modulation with the shortest modulation Period=$0^{d}.468002$ ever observed. The result detect small but definite modification in temperature and mean radius of the star itself. All results are compared with previously published literature values and discussed.
 \end{abstract}

{\footnotesize This paper was prepared with the \LaTeX \\}
\begin{multicols}%
{2}

\section{\normalsize Introduction} RV UMa is a RR Lyr type variable star (RRab type) in the constellation of Ursa Major, located (R.A. $13^\circ $33' 18.09'', Decl. +$53^\circ $59' 14.6''), and it has this parameter: \\ \\\textbf{V$_M$ range}: 9.81 - 11.3 \\\textbf{Spectral type}: A6-F5 \\ \textbf{Orbital Period}: P= $0^{d}.46806$. \\ \\RR Lyrae is a particular type of variables stars with asymmetric light curves (steep ascending branches), periods from 0.3 to 1.2 days, and amplitudes from 0.5 to 2 magnitude in V. This phenomenon of modulated light variation is called \textit{Blazhko effect}.
\section{\normalsize Data} \subsection{\normalsize Observations}
The stars was observed in 2011 April and May (UT) and all data were obtained with the Richey-Chretien telescope of the TS Corporation on Andrate (TO) - Italy station, equipped with a CCD camera (FLI EEV2 back illuminated, 2048×2048 pixel mm 0,39 arcsex/pix) with V filters (Johnson-Kron-Cousins \cite{Tasselli:2011ug}, as a good quality, homogeneous set with an accuracy of 0.01 - 0.02 mag. \\
Fig. 1 show the identification map for the stars RV UMa and Ref used in this study. \\Table 1 show the RV UMa data by C.D.S. - SIMBAD, Table 2 show the journal book of observations and capture image. We do not indicate the values of the Dark Frame as the CCD is cooled to liquid nitrogen, and therefore there were no shooting Dark. Preliminary processing of all CCD frames, to apply bias and flat field corrections, was alone with standard routines in the IRAF software package.\\ The magnitudes of stars in the Table 3, 4, 5, 6, 7, 8 and 9 are computed by fitting the position and scale of the PSF to each star image in turn, in order of decreasing brightness. The Zero Point of the frame is set during the PSF calculation, thought aperture photometry of the stars used to calculate the PSF.
\subsection{\normalsize Transformations and Reductions}
Instrumental magnitudes for all measured stars were transformed to a standard system using fitting coefficients derived from observations of standard stars whose magnitudes have been well established in earlier studies\cite{Tasselli:ref1}. \\The standard data for RV UMa and Ref used in this study, are visible in Table 1 from C.D.S. - SIMBAD\cite{(Catalogo:ref1}
\subsection{\normalsize Photometric Error}
To determine the error in the measurement of values used the method of the standard deviation. The following table highlights the values obtained in this study: \newline \\
\begin{tabular}{|c|c|}
\hline
\bf {\small Data} & \bf {\small Photometric Error} \\
\hline
{\small Apr 08} & {\small 0,63}\\
\hline
{\small Apr 13} &  {\small 1,41}\\
\hline
{\small May 05} & {\small -0,08} \\
\hline
{\small May 27} &  {\small -0,69}\\
\hline
\end{tabular}  \\ \\
{\textbf{{\scriptsize Photometric error for days}}} \\ \\
\normalsize Figures 3, 4, 5 and 6 show the deviation for each observation session for the V band.
\subsection{\normalsize Comparison with Previous Studies}
The result of this work study of RV UMa, is comparable with other paper study, and the data has been published in Table 10.
\section{\normalsize Data Analysis}
The analysis of the data and their calibration to the international system, are shown by this study in Table 3, 4, 5, 6, 7, 8 and 9. \\Column 1 give the number of observation, Column 2 give the Time of observation in JD, Column 3 give the V magnitudo of RV UMa measured for the time in column 2, Column 4, 5 and 6 give the V magnitude data photometry for Star Ref, measured of this study.\\
\subsection{\normalsize The Light Curve diagrams and Period Data}
In Fig. 2 we can show the Light Curve Diagram star for this work. Data analysis in this work allowed to obtain the following result for RV UMa: \\ \\
\begin{tabular}{|c|c|c|}
\hline
{\bf {{\scriptsize Frequence}}} & {\bf {\scriptsize Amplitude}} & {\bf {\scriptsize Phase}} \\
\hline
{\scriptsize \textbf{ 1.56302245E+01} } & {\scriptsize  \textbf{1.93041372} }& {\scriptsize \textbf{0.468002} }\\
\hline
\end{tabular} \newline \\ \\
The LightCurve Diagram for all days of study are show at the end of the paper. 
\subsection{{\normalsize The Residual Diagrams}}
The residual curve of all the data (Fig. 8) shows that the distortion is symmetrical to phase 0.468002 which is set to the phase of the middle of the rising branch. Thus the phase on the ascending branch when the visual luminosity of the star equals its time average value (at phase 0.46 in the figures) are closely related, with only some minute differences, to the onset of the H emission. \\In Figs. 8 the symmetrical modulation is centred exactly on this phase of the pulsation, indicating a connection between the origin of the modulation and that of the H emission.\\ At the end of this paper are show the singular LightCurve Diagram and Residual LightCurve Diagram for all day of this study.

\subsection{\normalsize Fourier Diagrams}
The Fourier spectra of the V data and the data prewhitened with the pulsation frequency and its harmonics are shown in Fig.7. At the end of the paper are show the singular Fourier diagrams for all day of this study.
\section{\normalsize Conclusion}
The study variable RV UMa, has highlighted the presence Blazhko effect especially highlighted in the diagram for the period of analysis, Fig. 2. The data obtained are also in line with the estimates contained in the data AAVSO.\cite{AAVSO:ref1} \\
The new results concerning the properties of the modulation of RV UMa are summarized in the next items:
\begin{itemize}
\item Residual scatter of the light curve is still concentrated in the ascending branch, indicating some irregular behaviour of the modulation;
\item The modulation is concentrated to 0.46 phase interval of the pulsation (Fig.8);
\item The symmetrical modulation centred exactly on this phase (0.468002) of the pulsation, indicating a connection between the origin of the modulation and that of the H emission. This is show in Fig.8
\end{itemize}
Give the importance of this type of stars and their important role in astrophysics not only as basic distance indicator, but also Blazhko effect and because as the most studied star in the pulsation and evolution of Pop II object, and know the still insufficient set of data available, further observations are needed.
\section{\normalsize Acknowledgments}
I would like to thank Dr.ssa Silvia Gargano for supporting and helping me during this study, with useful comments and helpful discussions which were extremely valuable. The constructive comments are highly appreciated.
\end{multicols}
\newpage
\centering
\newpage
\begin{table}
\caption{The RV UMa Data from C.D.S. - SIMBAD\cite{(Catalogo:ref1}} 
\small
\begin{tabular}{|c|c|c|c|c|c|c|c|}
\hline
{\bf Star} &    {\bf r} &    {\bf J} &    {\bf H} &    {\bf K} &    {\bf V} &   {\bf Rc} & {\bf V-Rc} \\
\hline
           &            &            &            &            &            &            &            \\
\hline
    RV UMa& -     & 10.08     & 9.83     & 9.81     & 10.9   &   - & -    \\
\hline
      Ref1 &   -   &  -   &  -  &   -   &  12.04    &  11.58    &  0.46     \\
\hline
      Ref2 &    -  & -    & -     & -     & 12.56    &   12.37   &   0.19  \\
\hline
      Ref3 &   -  & -    & -    &   -  &  12.98   & 12.43    &   0.55   \\
\hline
\end{tabular}  
\end{table}
\begin{table}
\tiny
\caption{Journal Book of observations and capture image } 
\tiny
\begin{tabular}{|r|r|r|r|r|r|r|r|cccc|}
\hline
\multicolumn{ 1}{|c}{{\bf Data}} & \multicolumn{ 1}{|c}{{\bf Sky}} & \multicolumn{ 1}{|c}{{\bf Seeing}} & \multicolumn{ 1}{|c}{{\bf Moon}} & \multicolumn{ 1}{|c}{{\bf Umidity}} & \multicolumn{ 1}{|c}{{\bf Filter}} & \multicolumn{ 1}{|c}{{\bf Exposition Time}} & \multicolumn{ 1}{|c}{{\bf Open Circles}} &           \multicolumn{ 4}{|c|}{{\bf Calibration}} \\
\multicolumn{ 1}{|c}{{\bf }} & \multicolumn{ 1}{|c}{} & \multicolumn{ 1}{|c}{} & \multicolumn{ 1}{|c}{} & \multicolumn{ 1}{|c}{} & \multicolumn{ 1}{|c}{{\bf }} & \multicolumn{ 1}{|c}{} & \multicolumn{ 1}{|c}{} & \multicolumn{ 2}{|c}{{\bf Dark}} & \multicolumn{ 2}{|c|}{{\bf Flat Field}} \\
\multicolumn{ 1}{|c}{} & \multicolumn{ 1}{|c}{} & \multicolumn{ 1}{|c|}{} & \multicolumn{ 1}{c}{} & \multicolumn{ 1}{|c}{} & \multicolumn{ 1}{|c}{} & \multicolumn{ 1}{|c}{} & \multicolumn{ 1}{|c|}{} & {\bf Number} & {\bf Time of exposition} & {\bf Number} & {\bf Time of exposition} \\
\hline
           &            &            &            &            &            &            &            &            &            &            &            \\
08/04/2011 &      Clear &       7/10 &       NO &       50\% &          V &    120 sec & 2, 3 and 7 &          - &          - &         30 &  120 sec     \\
13/04/2011 &      Clear &       3/5 &       YES &       70\% &          V &    120 sec & 2, 3 and 7 &          - &          - &         30 &    120 sec \\
05/05/2011 &      Clouds &       2/5 &       NO &       50\% &          V &    120 sec & 2, 3 and 7 &          - &          - &         30 &     120 sec \\
27/05/2011 &      Clear &       3/5 &       YES &       50\% &          V &    120 sec & 2, 3 and 7 &          - &          - &         30 &       120 sec \\
\hline
\end{tabular}  
\end{table}
\begin{table}
\tiny
\caption{V Magnitudo Data for RV UMa and Ref stars} 
\begin{tabular}{|c|c|c|c|c|c||c|c|c|c|c|c|}
\hline
  {\bf ID} & {\bf T (JD)} & {\bf Obj1} & {\bf Ref1} & {\bf Ref2} & {\bf Ref3}  &   {\bf ID} & {\bf T (JD)} & {\bf Obj1} & {\bf Ref1} & {\bf Ref2} & {\bf Ref3}  \\
\hline
         1 & 2455660.491 &      11.93 &       0.49 &       0.50 &       0.48 &         86 & 2455660.646 &       8.62 &       0.50 &       0.51 &       0.49 \\
\hline
         2 & 2455660.493 &      11.81 &       0.48 &       0.49 &       0.47 &         87 & 2455660.648 &       8.66 &       0.49 &       0.50 &       0.48 \\
\hline
         3 & 2455660.495 &      11.72 &       0.49 &       0.50 &       0.48 &         88 & 2455660.651 &       8.57 &       0.49 &       0.50 &       0.48 \\
\hline
         4 & 2455660.496 &      11.78 &       0.48 &       0.49 &       0.47 &         89 & 2455660.652 &       8.62 &       0.49 &       0.50 &       0.48 \\
\hline
         5 & 2455660.498 &      11.67 &       0.48 &       0.49 &       0.47 &         90 & 2455660.654 &       8.60 &       0.49 &       0.50 &       0.48 \\
\hline
         6 &  2455660.5 &      11.63 &       0.49 &       0.50 &       0.48 &         91 & 2455660.656 &       8.57 &       0.47 &       0.48 &       0.46 \\
\hline
         7 & 2455660.501 &      11.55 &       0.49 &       0.50 &       0.48 &         92 & 2455660.657 &       8.59 &       0.48 &       0.49 &       0.47 \\
\hline
\end{tabular}  
\end{table}
\begin{table}
\tiny
\caption{V Magnitudo Data for RV UMa and Ref stars} 
\begin{tabular}{|c|c|c|c|c|c||c|c|c|c|c|c|}
\hline
  {\bf ID} & {\bf T (JD)} & {\bf Obj1} & {\bf Ref1} & {\bf Ref2} & {\bf Ref3}  &   {\bf ID} & {\bf T (JD)} & {\bf Obj1} & {\bf Ref1} & {\bf Ref2} & {\bf Ref3} \\
\hline
176 & 24556653230358900 &      12.44 &       0.49 &       0.50 &       0.48 &        238 & 24556653584868100 &      12.76 &       0.47 &       0.48 &       0.46 \\
\hline
       177 & 24556653235921600 &      12.53 &       0.49 &       0.50 &       0.48 &        239 & 24556653591496900 &      12.83 &       0.48 &       0.49 &       0.47 \\
\hline
       178 & 24556653241480500 &      12.51 &       0.49 &       0.50 &       0.48 &        240 & 24556653597089400 &      12.81 &       0.50 &       0.51 &       0.49 \\
\hline
       179 & 24556653247022100 &      12.39 &       0.49 &       0.50 &       0.48 &        241 & 24556653602708600 &      12.82 &       0.48 &       0.49 &       0.47 \\
\hline
       180 & 24556653253416600 &      12.48 &       0.48 &       0.49 &       0.47 &        242 & 24556653608257100 &      12.83 &       0.46 &       0.47 &       0.45 \\
\hline
       181 & 24556653258990600 &      12.60 &       0.48 &       0.49 &       0.47 &        243 & 24556653613801000 &      12.90 &       0.49 &       0.50 &       0.48 \\
\hline
       182 & 24556653265331900 &      12.41 &       0.49 &       0.50 &       0.48 &        244 & 24556653619799300 &      12.73 &       0.51 &       0.52 &       0.50 \\
\hline
\end{tabular}  
\end{table}
\begin{table}
\tiny
\caption{V Magnitudo Data for RV UMa and Ref stars} 
\begin{tabular}{|c|c|c|c|c|c||c|c|c|c|c|c|}
\hline
  {\bf ID} & {\bf T (JD)} & {\bf Obj1} & {\bf Ref1} & {\bf Ref2} & {\bf Ref3}  &   {\bf ID} & {\bf T (JD)} & {\bf Obj1} & {\bf Ref1} & {\bf Ref2} & {\bf Ref3} \\
\hline
 295 & 24556653916570000 &      13.20 &       0.47 &       0.48 &       0.46 &        357 & 24556654276419000 &      13.55 &       0.46 &       0.47 &       0.45 \\
\hline
       296 & 24556653922152100 &      13.07 &       0.48 &       0.49 &       0.47 &        358 & 24556654281996900 &      13.60 &       0.47 &       0.48 &       0.46 \\
\hline
       297 & 24556653927842000 &      13.16 &       0.48 &       0.49 &       0.47 &        359 & 24556654288781000 &      13.54 &       0.47 &       0.48 &       0.46 \\
\hline
       298 & 24556653933438700 &      13.21 &       0.49 &       0.50 &       0.48 &        360 & 24556654294336400 &      13.48 &       0.48 &       0.49 &       0.47 \\
\hline
       299 & 24556653939086900 &      13.19 &       0.49 &       0.50 &       0.48 &        361 & 24556654299909200 &      13.53 &       0.48 &       0.49 &       0.47 \\
\hline
       300 & 24556653944666700 &      13.19 &       0.49 &       0.50 &       0.48 &        362 & 24556654305528400 &      13.46 &       0.48 &       0.49 &       0.47 \\
\hline
       301 & 24556653950225600 &      13.16 &       0.48 &       0.49 &       0.47 &        363 & 24556654311118700 &      13.48 &       0.47 &       0.48 &       0.46 \\
\hline
\end{tabular}  
\end{table}
\begin{table}
\tiny
\caption{V Magnitudo Data for RV UMa and Ref stars} 
\begin{tabular}{|c|c|c|c|c|c||c|c|c|c|c|c|}
\hline
  {\bf ID} & {\bf T (JD)} & {\bf Obj1} & {\bf Ref1} & {\bf Ref2} & {\bf Ref3}  &   {\bf ID} & {\bf T (JD)} & {\bf Obj1} & {\bf Ref1} & {\bf Ref2} & {\bf Ref3} \\
\hline
       419 & 24556654632343000 &      13.73 &       0.47 &       0.48 &       0.46 &        481 & 24556654984791000 &      13.73 &       0.46 &       0.47 &       0.45 \\
\hline
       420 & 24556654637922800 &      13.81 &       0.48 &       0.49 &       0.47 &        482 & 24556654990333600 &      13.85 &       0.47 &       0.48 &       0.46 \\
\hline
       421 & 24556654643471200 &      13.87 &       0.48 &       0.49 &       0.47 &        483 & 24556654995913500 &      13.81 &       0.47 &       0.48 &       0.46 \\
\hline
       422 & 24556654649026700 &      13.82 &       0.49 &       0.50 &       0.48 &        484 & 24556655001519900 &      13.85 &       0.48 &       0.49 &       0.47 \\
\hline
       423 & 24556654654640100 &      13.81 &       0.49 &       0.50 &       0.48 &        485 & 24556655008151500 &      13.84 &       0.48 &       0.49 &       0.47 \\
\hline
       424 & 24556654660185100 &      13.77 &       0.49 &       0.50 &       0.48 &        486 & 24556655014564600 &      13.83 &       0.48 &       0.49 &       0.47 \\
\hline
       425 & 24556654666819600 &      13.77 &       0.48 &       0.49 &       0.47 &        487 & 24556655020131700 &      13.79 &       0.47 &       0.48 &       0.46 \\
\hline
\end{tabular}  
\end{table}
\begin{table}
\tiny
\caption{V Magnitudo Data for RV UMa and Ref stars} 
\begin{tabular}{|c|c|c|c|c|c||c|c|c|c|c|c|}
\hline
  {\bf ID} & {\bf T (JD)} & {\bf Obj1} & {\bf Ref1} & {\bf Ref2} & {\bf Ref3}  &   {\bf ID} & {\bf T (JD)} & {\bf Obj1} & {\bf Ref1} & {\bf Ref2} & {\bf Ref3} \\
\hline
       543 & 24556655339472800 &      13.70 &       0.47 &       0.48 &       0.46 &        605 & 24556655718554200 &      13.47 &       0.46 &       0.47 &       0.45 \\
\hline
       544 & 24556655344773200 &      13.69 &       0.48 &       0.49 &       0.47 &        606 & 24556655724989400 &      13.46 &       0.47 &       0.48 &       0.46 \\
\hline
       545 & 24556655350096800 &      13.69 &       0.48 &       0.49 &       0.47 &        607 & 24556655730584300 &      13.48 &       0.47 &       0.48 &       0.46 \\
\hline
       546 & 24556655355420400 &      13.76 &       0.49 &       0.50 &       0.48 &        608 & 24556655737295300 &      13.30 &       0.48 &       0.49 &       0.47 \\
\hline
       547 & 24556655360746200 &      13.59 &       0.49 &       0.50 &       0.48 &        609 & 24556655742928700 &      13.38 &       0.48 &       0.49 &       0.47 \\
\hline
       548 & 24556655366071000 &      13.81 &       0.49 &       0.50 &       0.48 &        610 & 24556655748478300 &      13.40 &       0.48 &       0.49 &       0.47 \\
\hline
       549 & 24556655394798600 &      13.79 &       0.48 &       0.49 &       0.47 &        611 & 24556655754055900 &      13.35 &       0.47 &       0.48 &       0.46 \\
\hline
\end{tabular}  
\end{table}
\begin{table}
\tiny
\caption{V Magnitudo Data for RV UMa and Ref stars} 
\begin{tabular}{|c|c|c|c|c|c||c|c|c|c|c|c|}
\hline
  {\bf ID} & {\bf T (JD)} & {\bf Obj1} & {\bf Ref1} & {\bf Ref2} & {\bf Ref3}  &   {\bf ID} & {\bf T (JD)} & {\bf Obj1} & {\bf Ref1} & {\bf Ref2} & {\bf Ref3} \\
\hline
       667 & 2455687.421 &      11.12 &      12.05 &      12.57 &      12.98 &        729 & 2455687.533 &      10.40 &      12.04 &      12.58 &      12.98 \\
\hline
       668 & 2455687.423 &      11.11 &      12.04 &      12.58 &      12.98 &        730 & 2455687.534 &      10.41 &      12.04 &      12.57 &      13.00 \\
\hline
       669 & 2455687.424 &      11.12 &      12.04 &      12.58 &      12.97 &        731 & 2455687.536 &      10.41 &      12.05 &      12.58 &      12.97 \\
\hline
       670 & 2455687.426 &      11.12 &      12.05 &      12.57 &      12.98 &        732 & 2455687.538 &      10.41 &      12.06 &      12.58 &      12.95 \\
\hline
       671 & 2455687.428 &      11.11 &      12.05 &      12.56 &      12.99 &        733 & 2455687.539 &      10.44 &      12.05 &      12.57 &      12.97 \\
\hline
       672 & 2455687.429 &      11.09 &      12.05 &      12.58 &      12.97 &        734 & 2455687.541 &      10.44 &      12.04 &      12.58 &      12.98 \\
\hline
       673 & 2455687.431 &      11.09 &      12.05 &      12.56 &      12.99 &        735 & 2455687.543 &      10.45 &      12.05 &      12.57 &      12.98 \\
\hline
\end{tabular}  
\end{table}
\begin{table}
\tiny
\caption{V Magnitudo Data for RV UMa and Ref stars} 
\begin{tabular}{|c|c|c|c|c|c||c|c|c|c|c|c|}
\hline
  {\bf ID} & {\bf T (JD)} & {\bf Obj1} & {\bf Ref1} & {\bf Ref2} & {\bf Ref3}  &   {\bf ID} & {\bf T (JD)} & {\bf Obj1} & {\bf Ref1} & {\bf Ref2} & {\bf Ref3} \\
\hline
       791 & 2455709.48 &      10.43 &      12.04 &      12.57 &      13.00 &            &            &            &            &            &            \\
\hline
       792 & 2455709.482 &      10.43 &      12.05 &      12.56 &      12.99 &            &            &            &            &            &            \\
\hline
       793 & 2455709.484 &      10.43 &      12.04 &      12.56 &      13.01 &            &            &            &            &            &            \\
\hline
       794 & 2455709.485 &      10.42 &      12.04 &      12.56 &      13.02 &            &            &            &            &            &            \\
\hline
       795 & 2455709.487 &      10.42 &      12.04 &      12.56 &      13.01 &            &            &            &            &            &            \\
\hline
       796 & 2455709.488 &      10.40 &      12.04 &      12.57 &      12.99 &            &            &            &            &            &            \\
\hline
       797 & 2455709.49 &      10.41 &      12.04 &      12.57 &      13.01 &            &            &            &            &            &            \\
\hline
\end{tabular}  
\end{table}
\begin{table}
\caption{Comparison of RV UMa parameters with published data} 
\begin{tabular}{|cccccr|}
\hline
{\bf Star} & {\bf Type} &             \multicolumn{ 4}{c|}{{\bf Period (d)}} \\
\hline
           &            &            &            &            &            \\
\hline
\multicolumn{ 1}{|c}{{\bf RV UMa}} & \multicolumn{ 1}{c}{{\bf RRAB}} & {\bf This Work} & {\bf GCVS} & {\bf A.V.Solov'ev, Tadj}  & {\bf AAVSSO} \\
\multicolumn{ 1}{|c}{{\bf }} & \multicolumn{ 1}{c}{{\bf }} &   0.468002 &   0.4680618 &  0.468062  &  0.46806 \\
\hline
\end{tabular}  
\end{table}
\clearpage
\newpage
\includegraphics{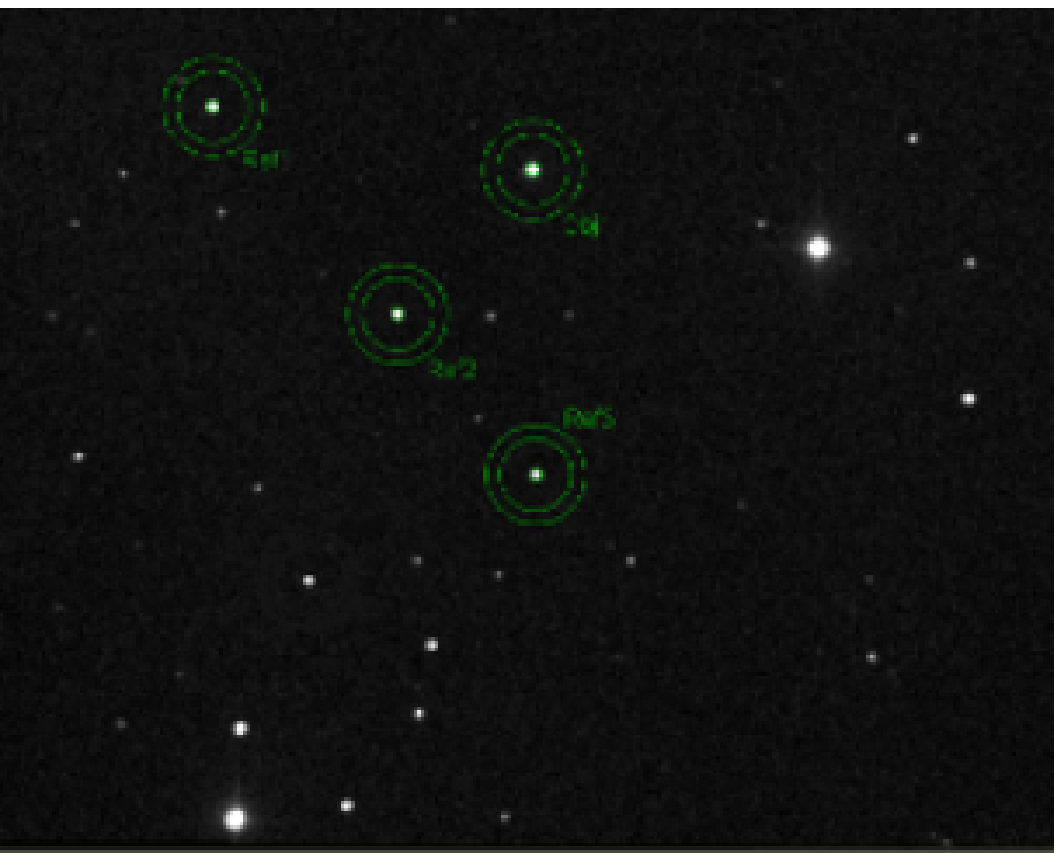} \\
\textbf{{\scriptsize Fig.1 Map of the stars RV UMa and Ref used in this paper}} \\
\includegraphics[width=0.5\textwidth{}]{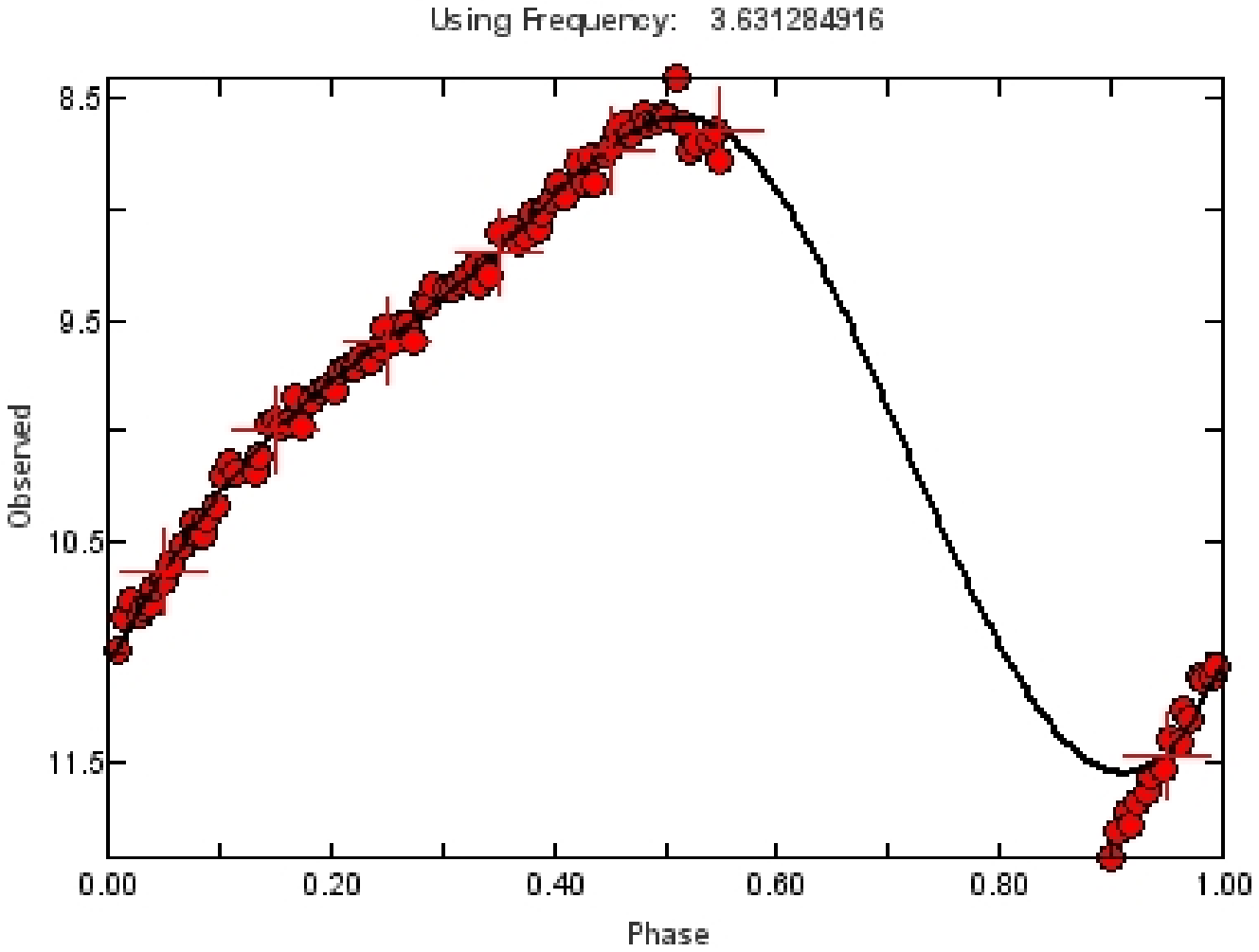} \\
\textbf{{\scriptsize Fig 2: V Light Curve Diagram of RV UMa for this study. The scatter of the light curve is caused by Blazhko modulation. Phase 0.468002 is set to the middle of the ascending branch defined as the phase where the V flux is equal to its time averaged value.}} \\ 
\includegraphics[width=0.5\textwidth{}]{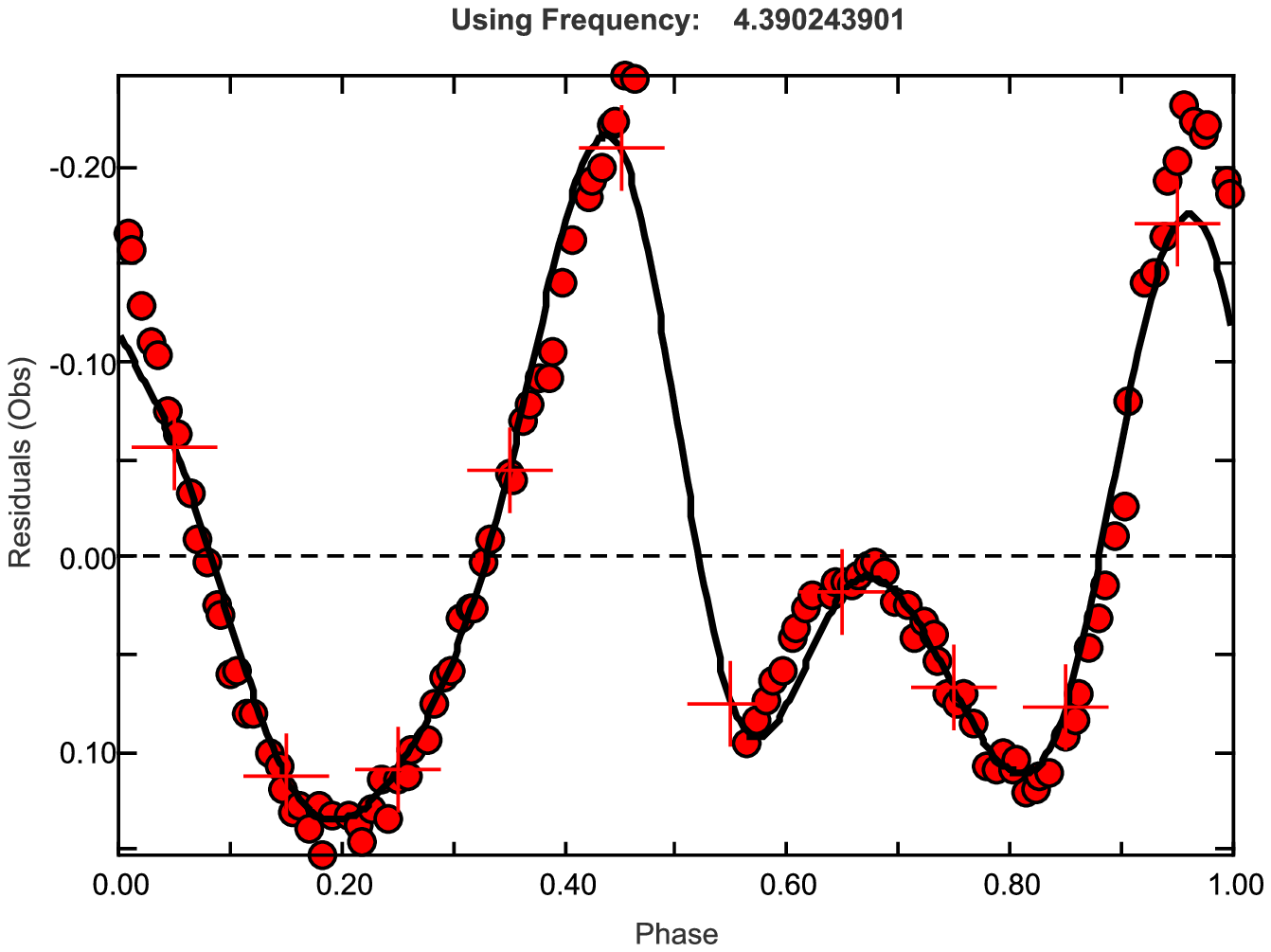} \\
\textbf{{\scriptsize Fig 8: Residual Light Curve Diagram for this study}} \\
\includegraphics[width=0.5\textwidth{}]{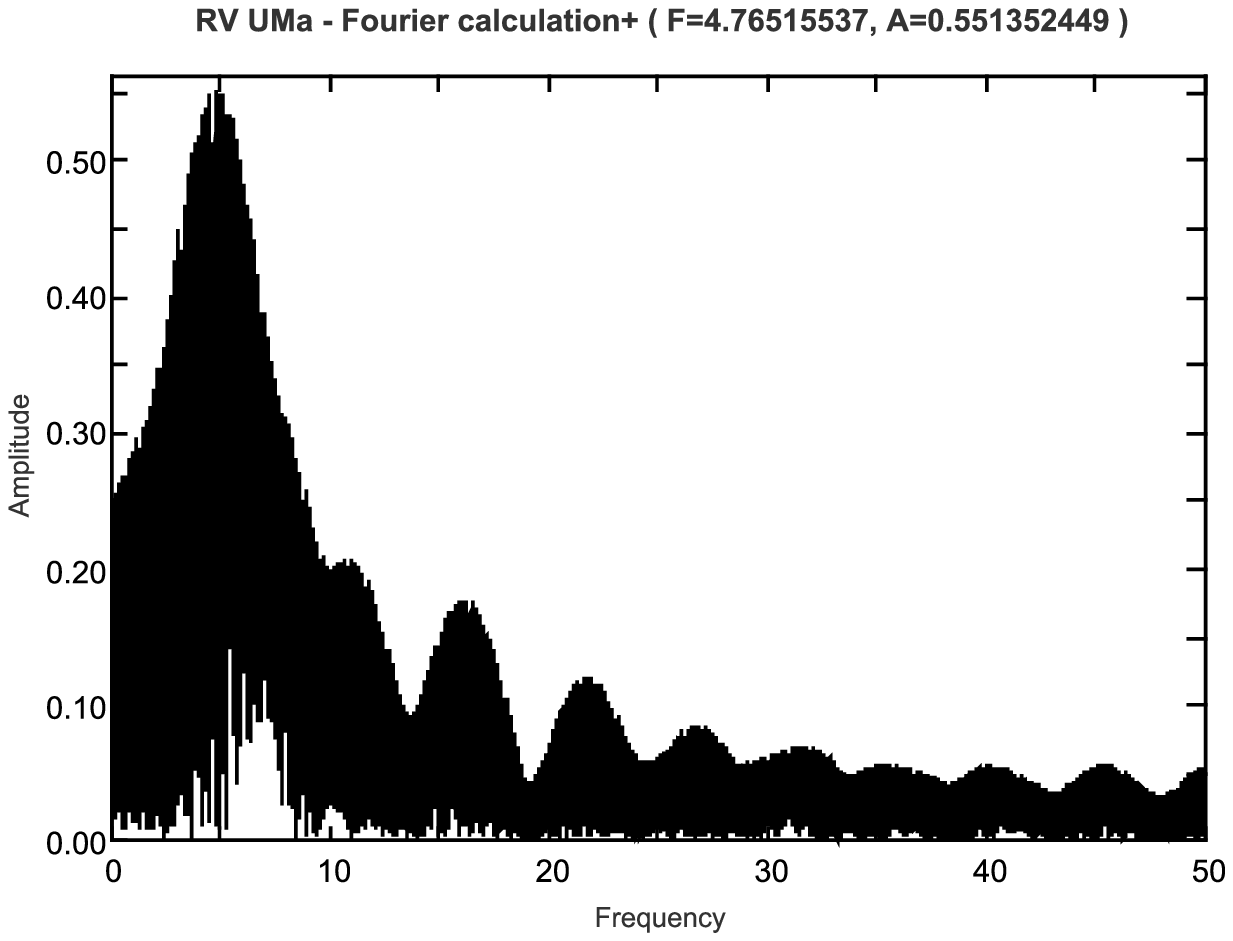}\\
\textbf{{\scriptsize Fig 7: Fourier Diagram of the V light curve of RV UMa for this study.}}\\
\includegraphics[width=0.48\textwidth{}]{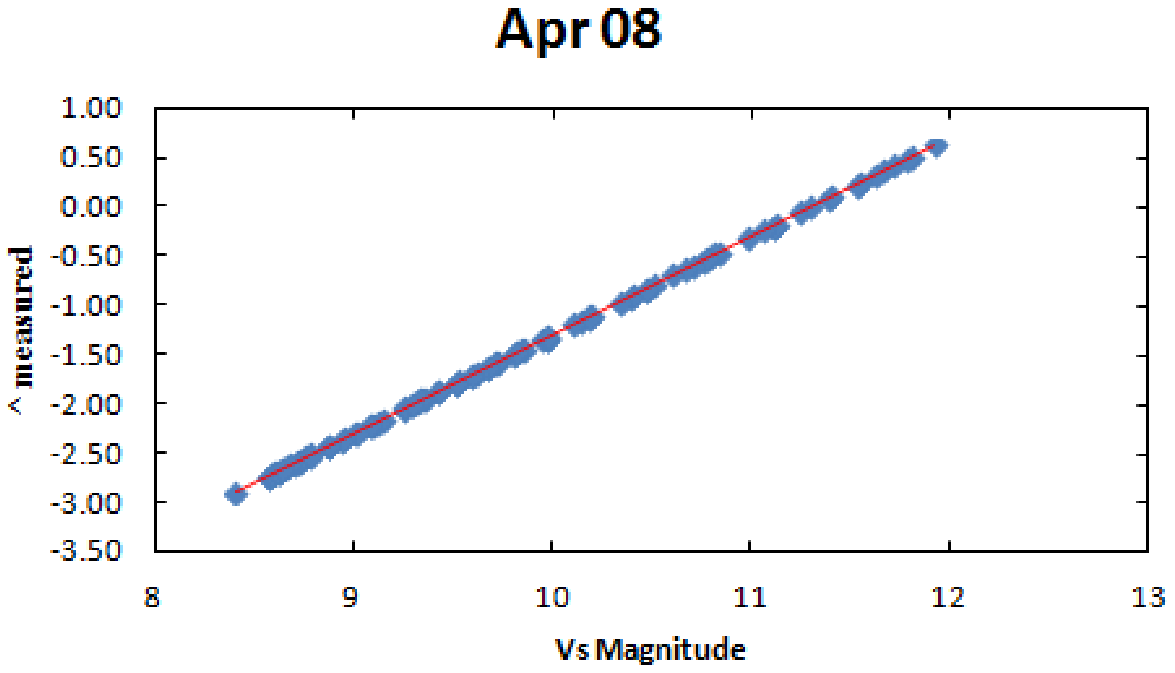} \quad 
\includegraphics[width=0.48\textwidth{}]{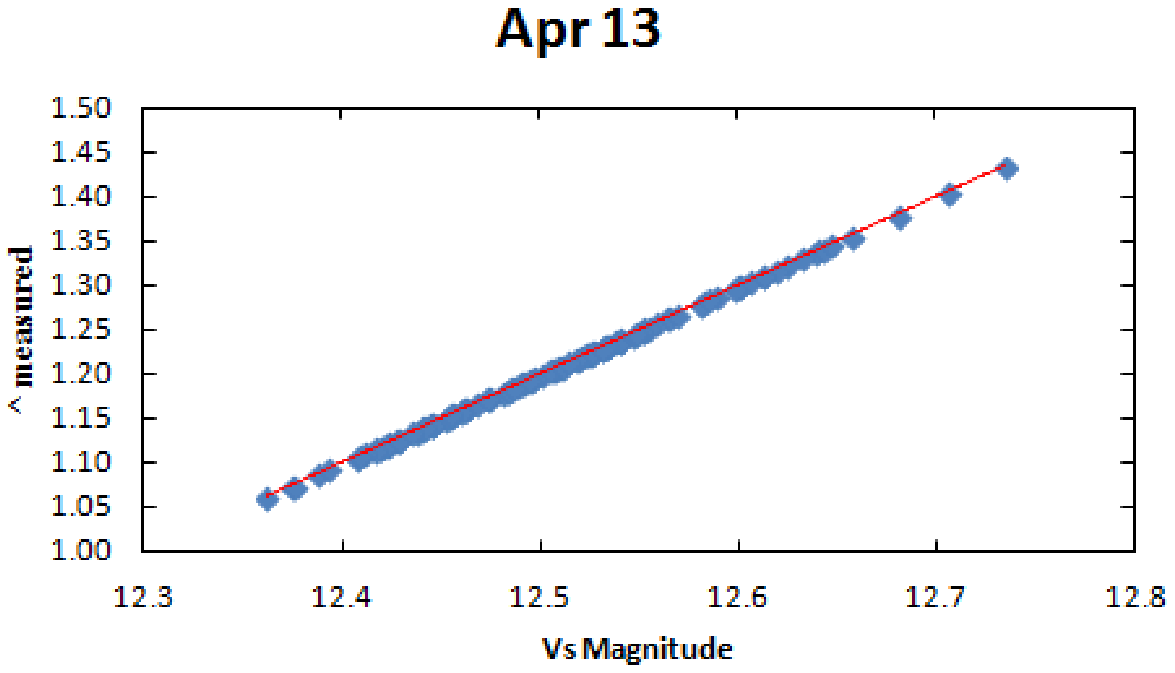}
\includegraphics[width=0.48\textwidth{}]{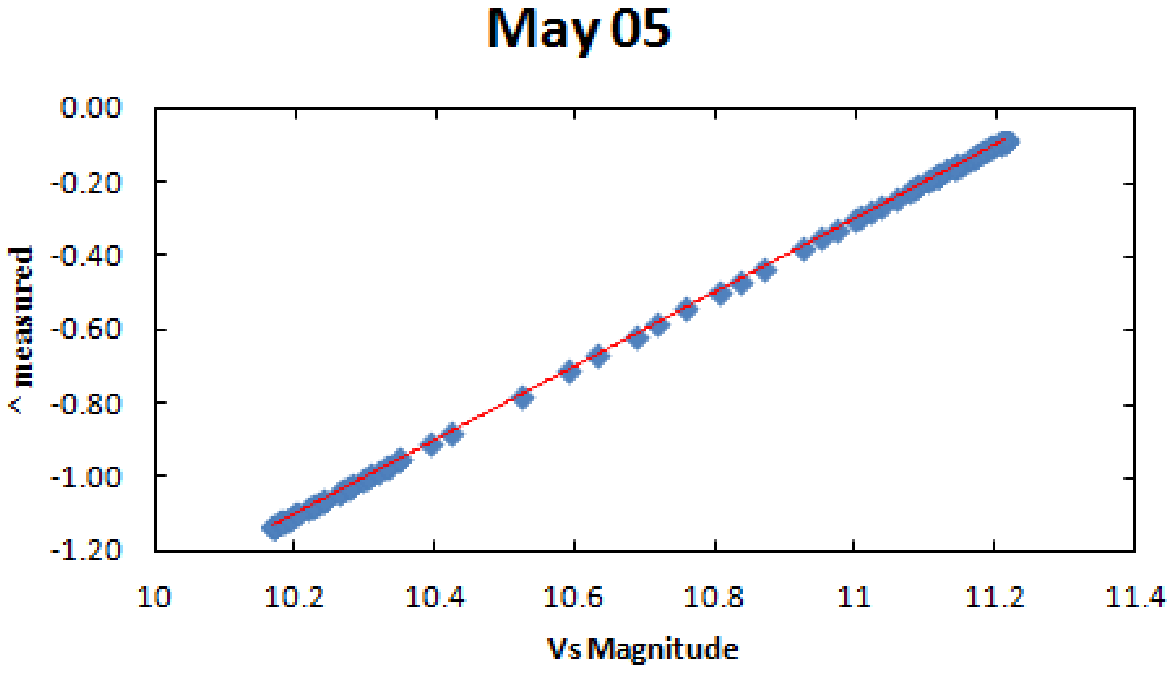} \quad
\includegraphics[width=0.48\textwidth{}]{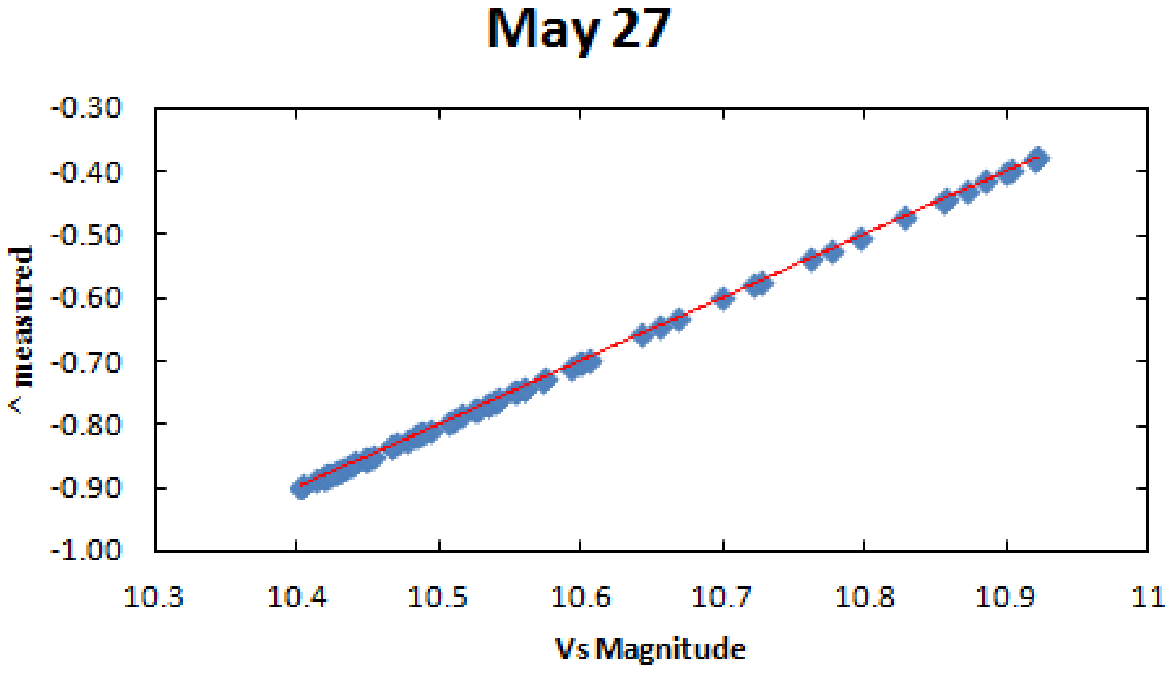} \\
\textbf{{\scriptsize Figure 3-4-5-6: variation between instrumental measurement and value stellar catalog for each observing session}} \newpage
\includegraphics[width=0.48\textwidth{}]{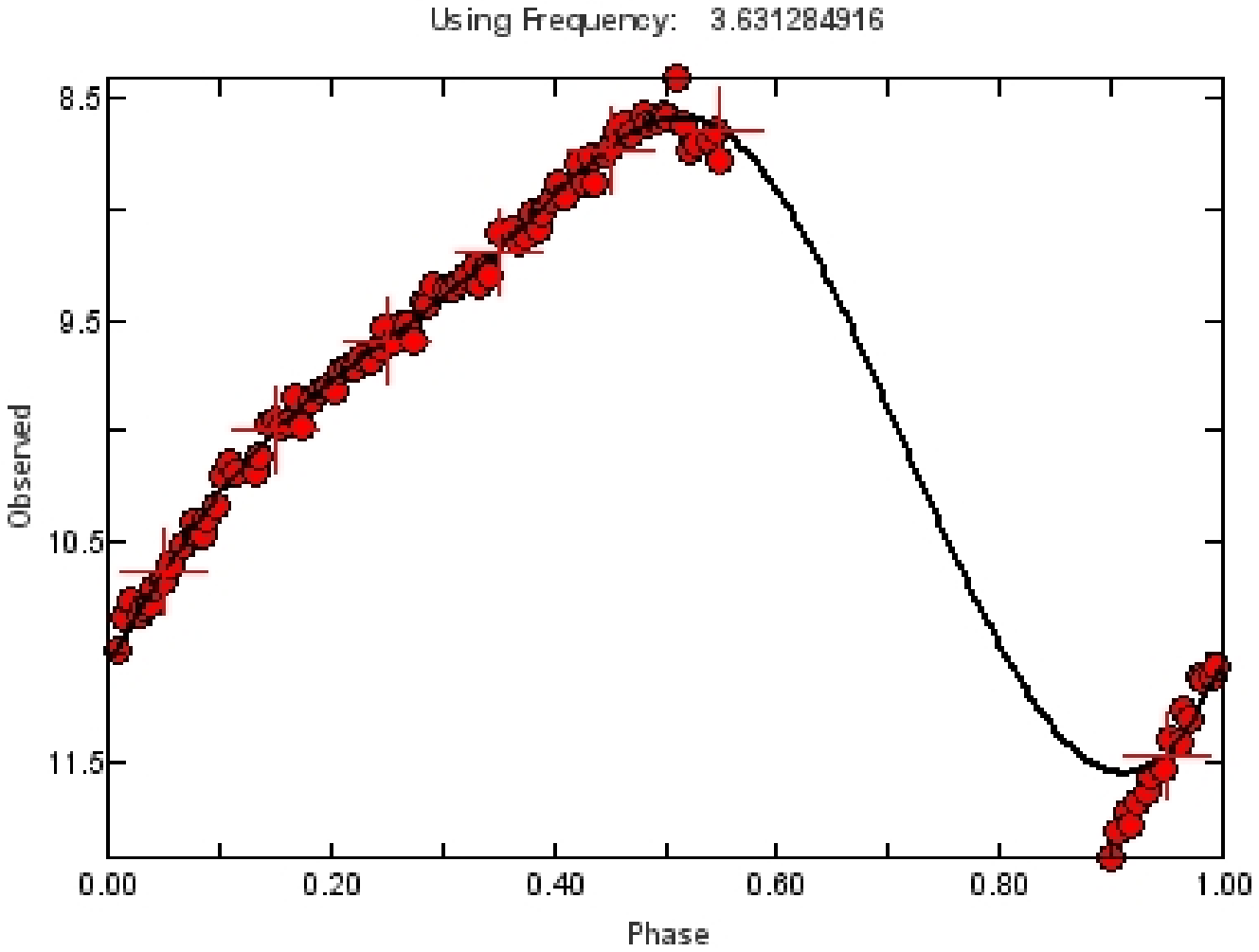}\quad
\includegraphics[width=0.48\textwidth{}]{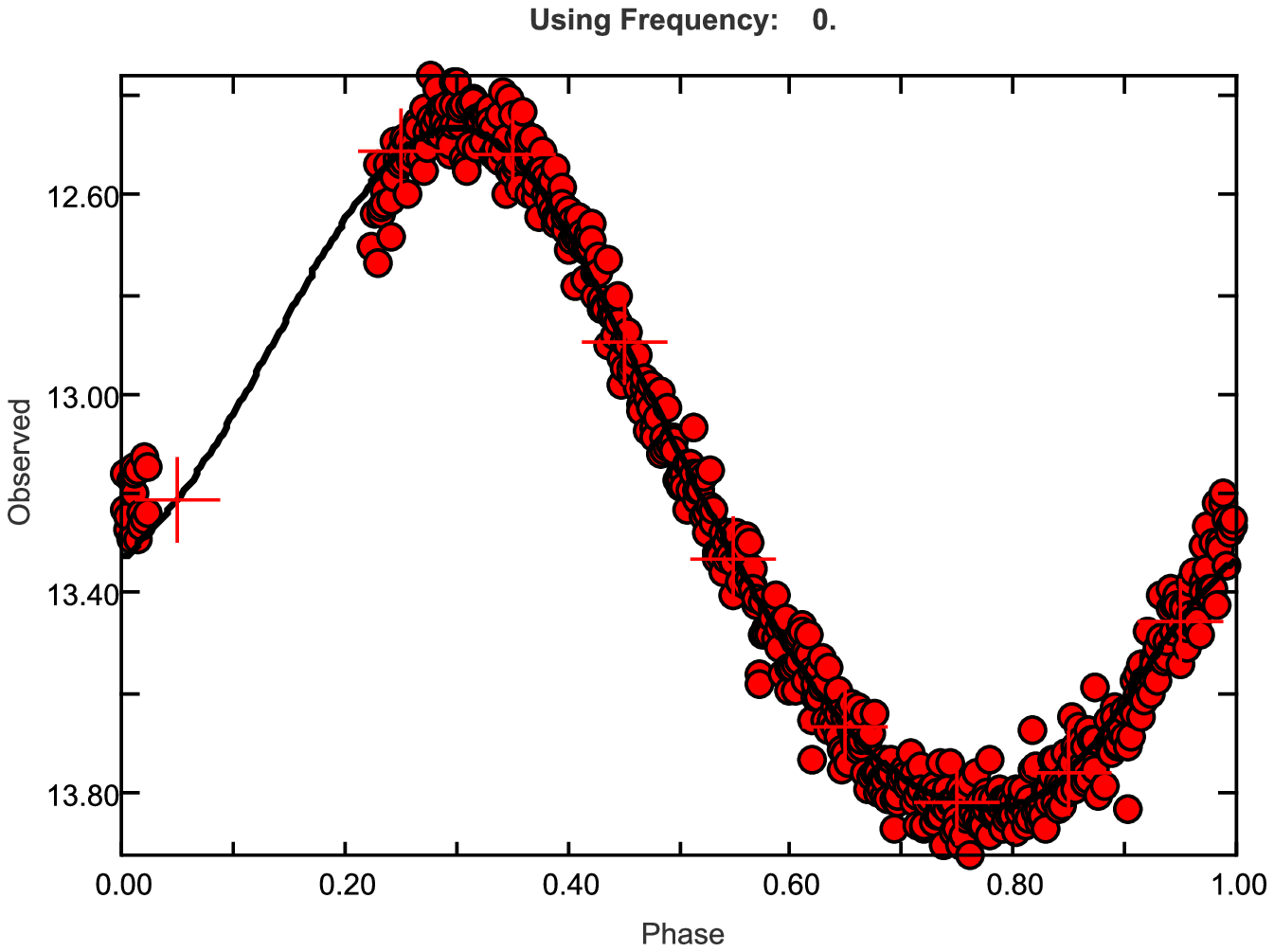}\\
\textbf{{\scriptsize Left: Lightcurve for Apr 8, 2011 - Right: Lightcurve for Apr 13, 2011}} 
\includegraphics[width=0.48\textwidth{}]{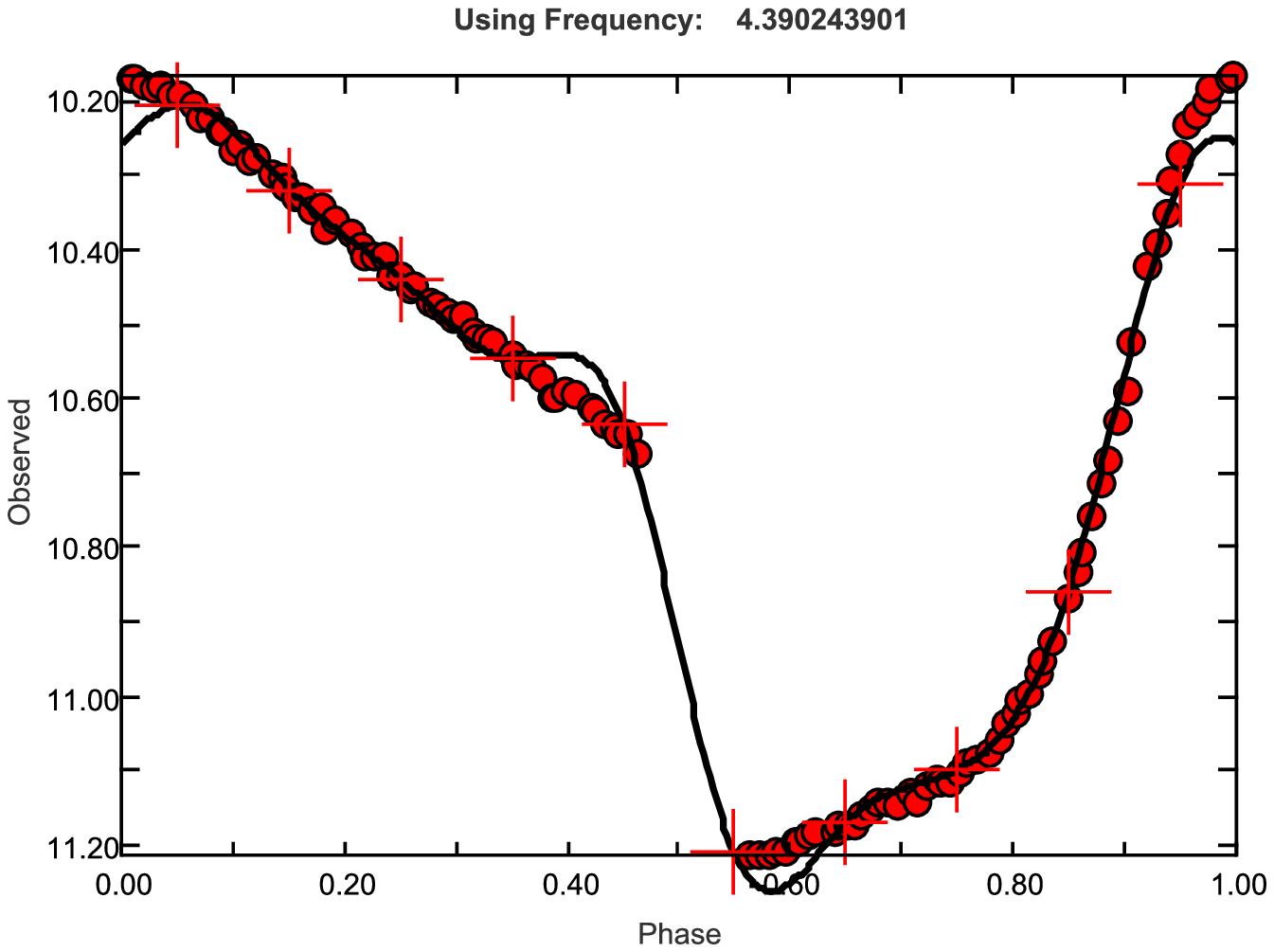}\quad
\includegraphics[width=0.48\textwidth{}]{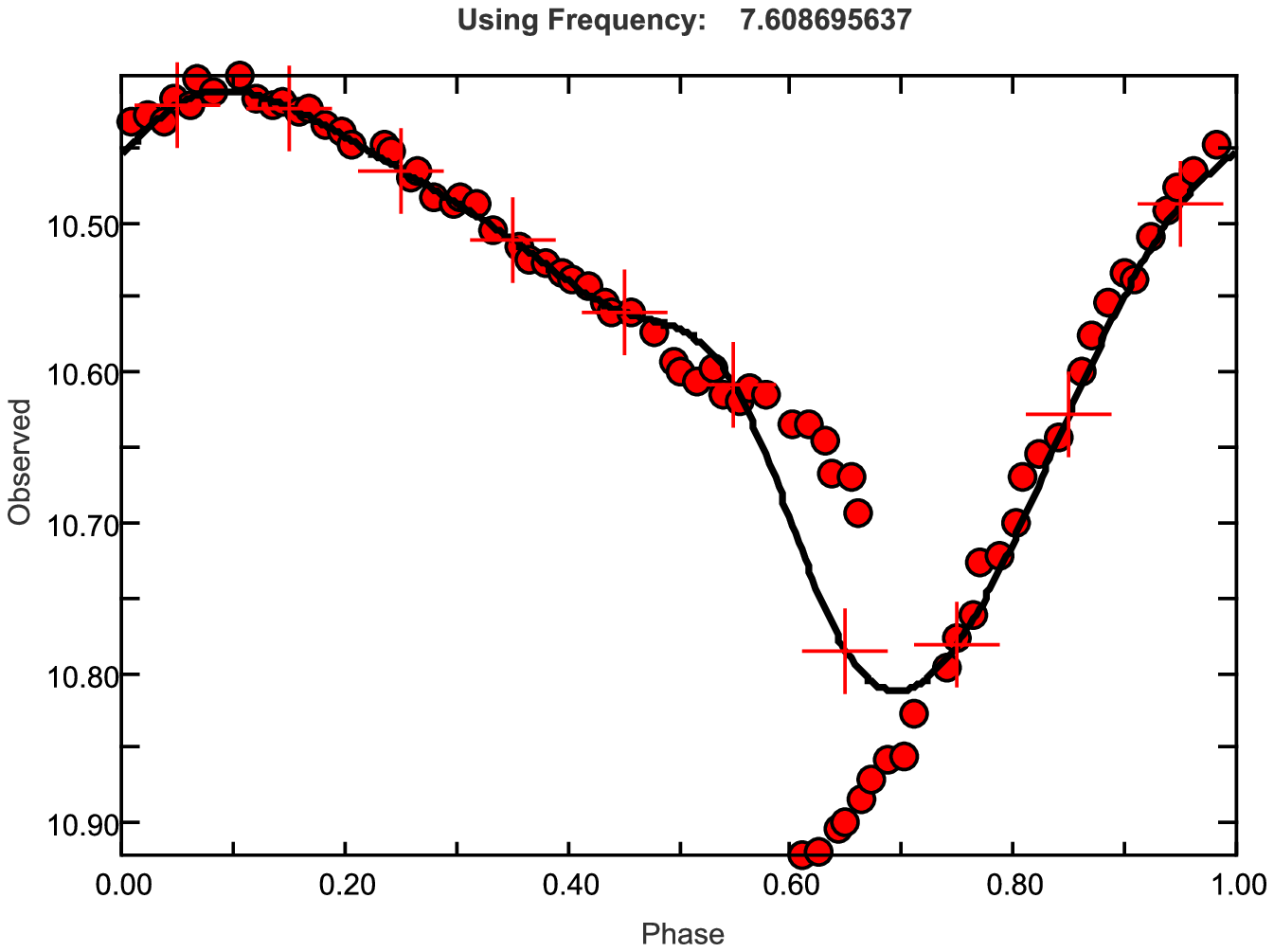}\\
\textbf{{\scriptsize Left: Lightcurve for May 5, 2011 - Right: Lightcurve for May 27, 2011}} \newpage
\includegraphics[width=0.48\textwidth{}]{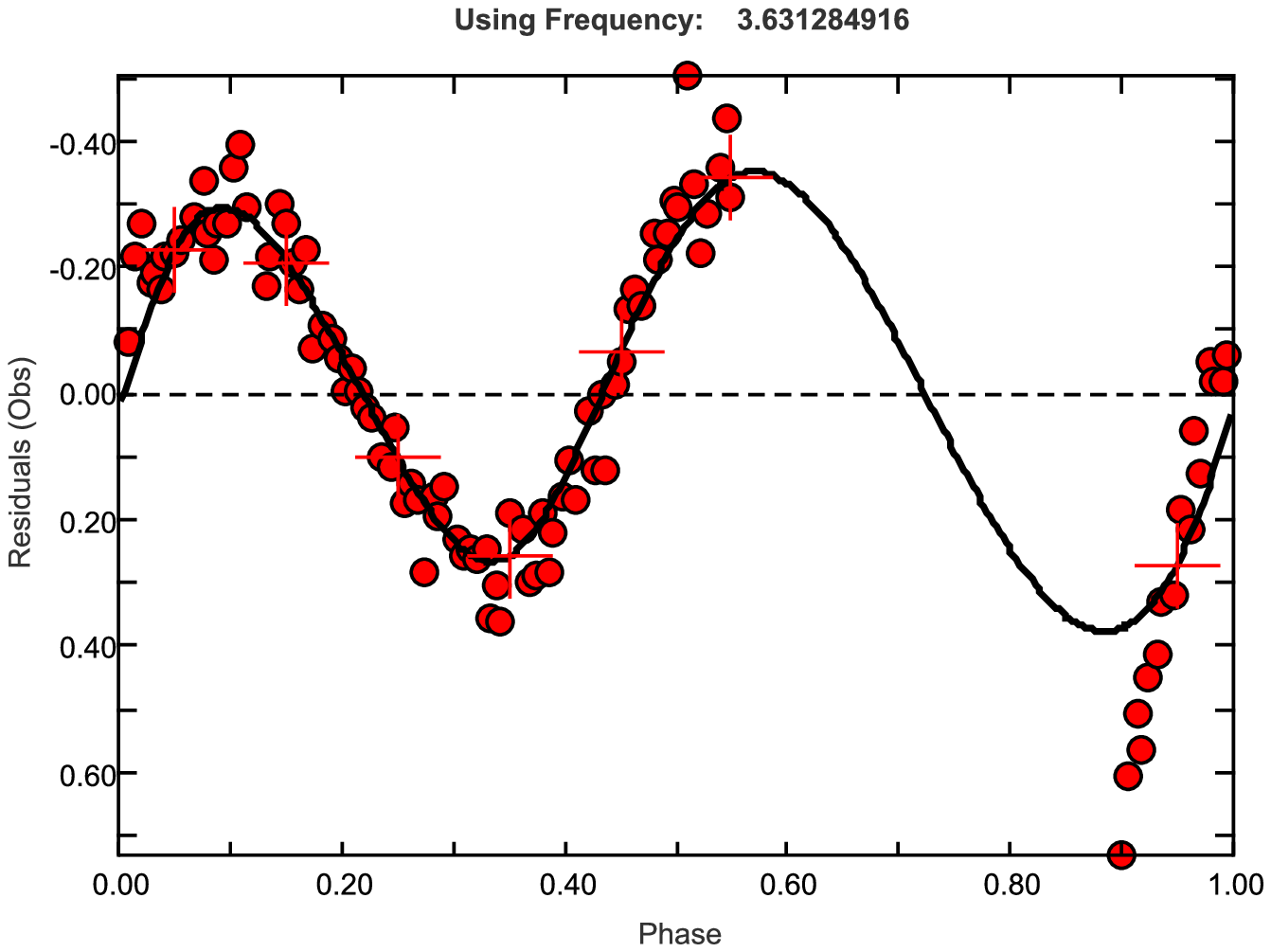}\quad
\includegraphics[width=0.48\textwidth{}]{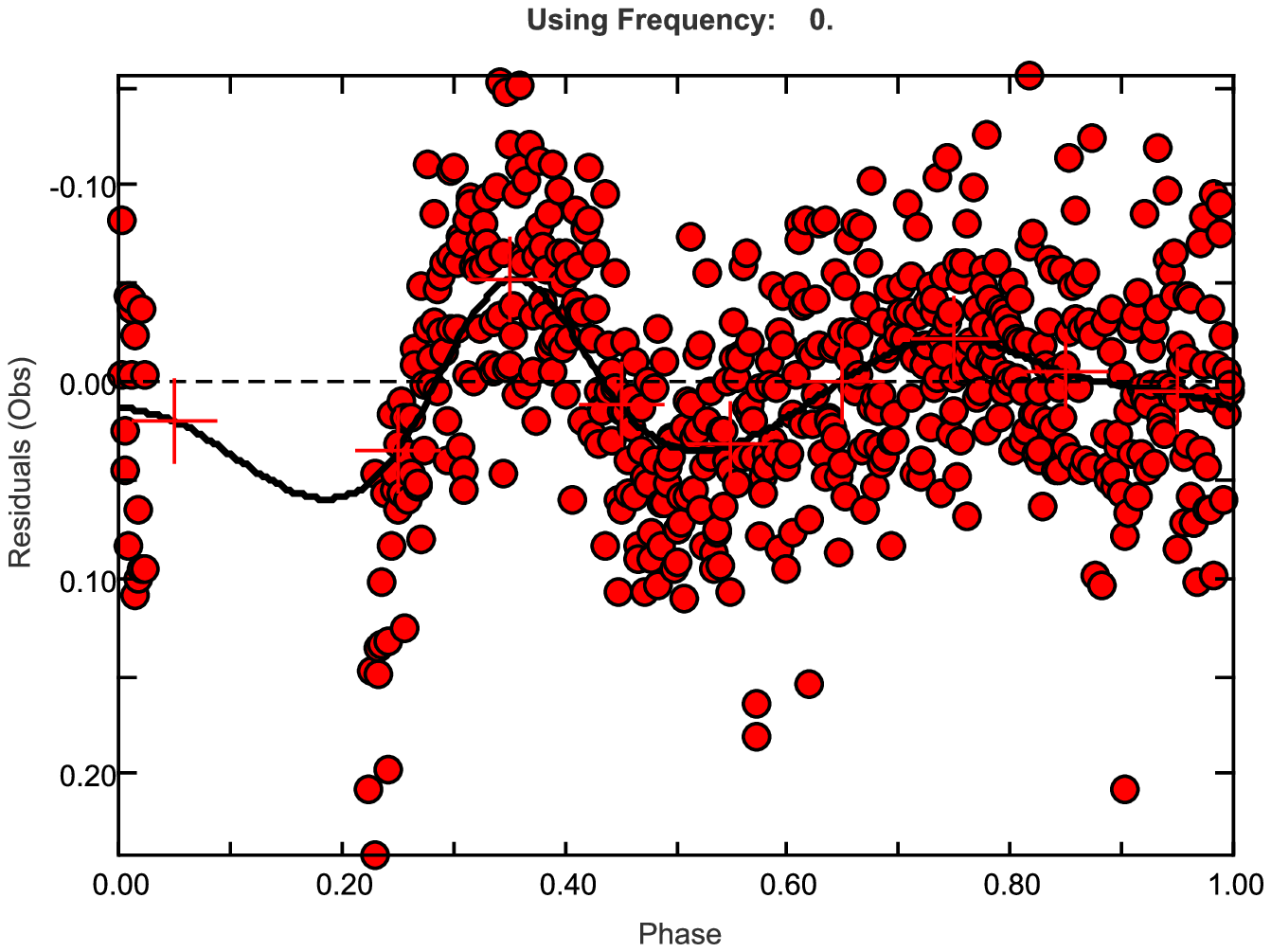}\\
\textbf{{\scriptsize Left: Residual for Apr 8, 2011 - Right: Residual for Apr 13, 2011}} 
\includegraphics[width=0.48\textwidth{}]{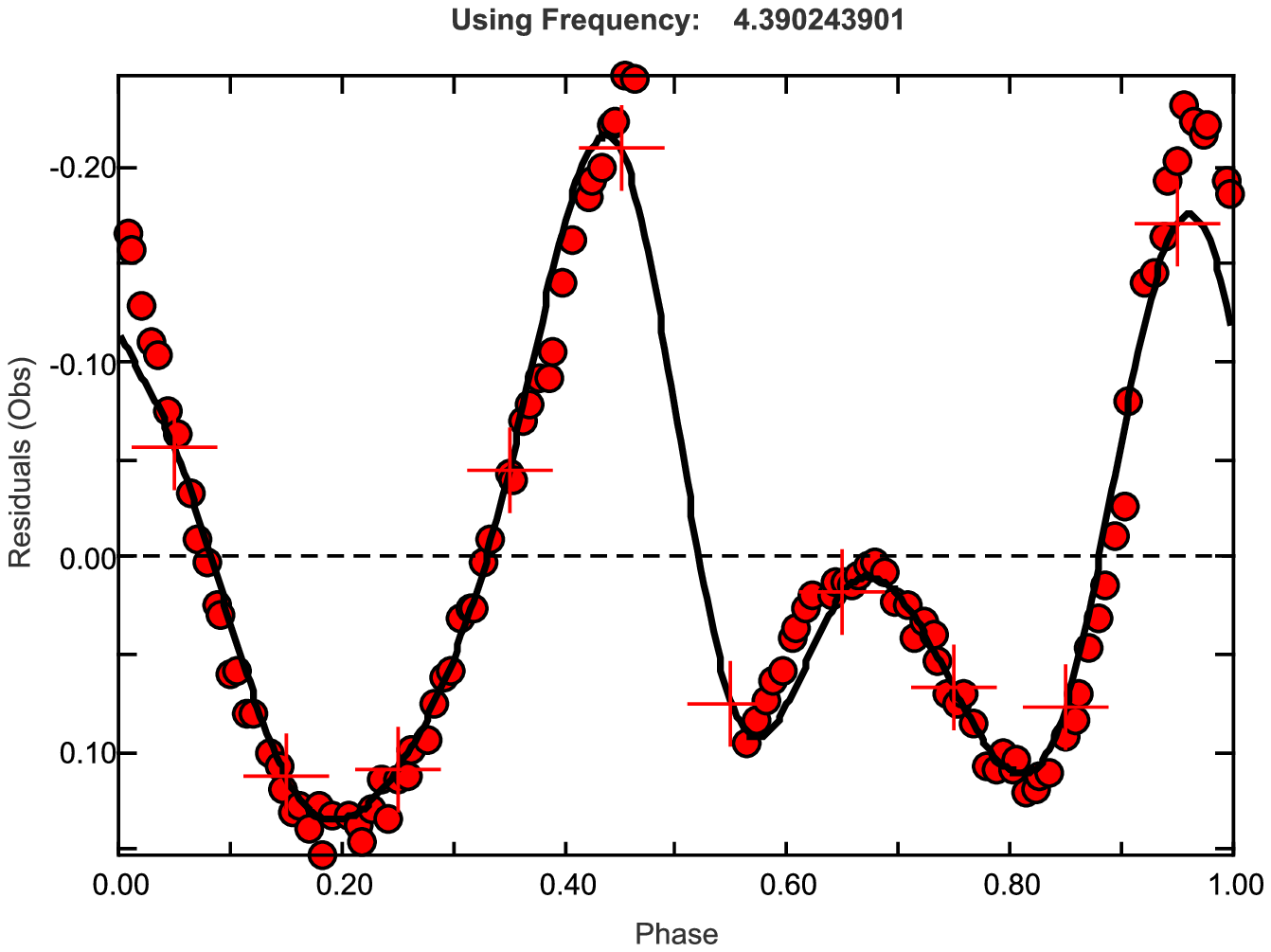}\quad
\includegraphics[width=0.48\textwidth{}]{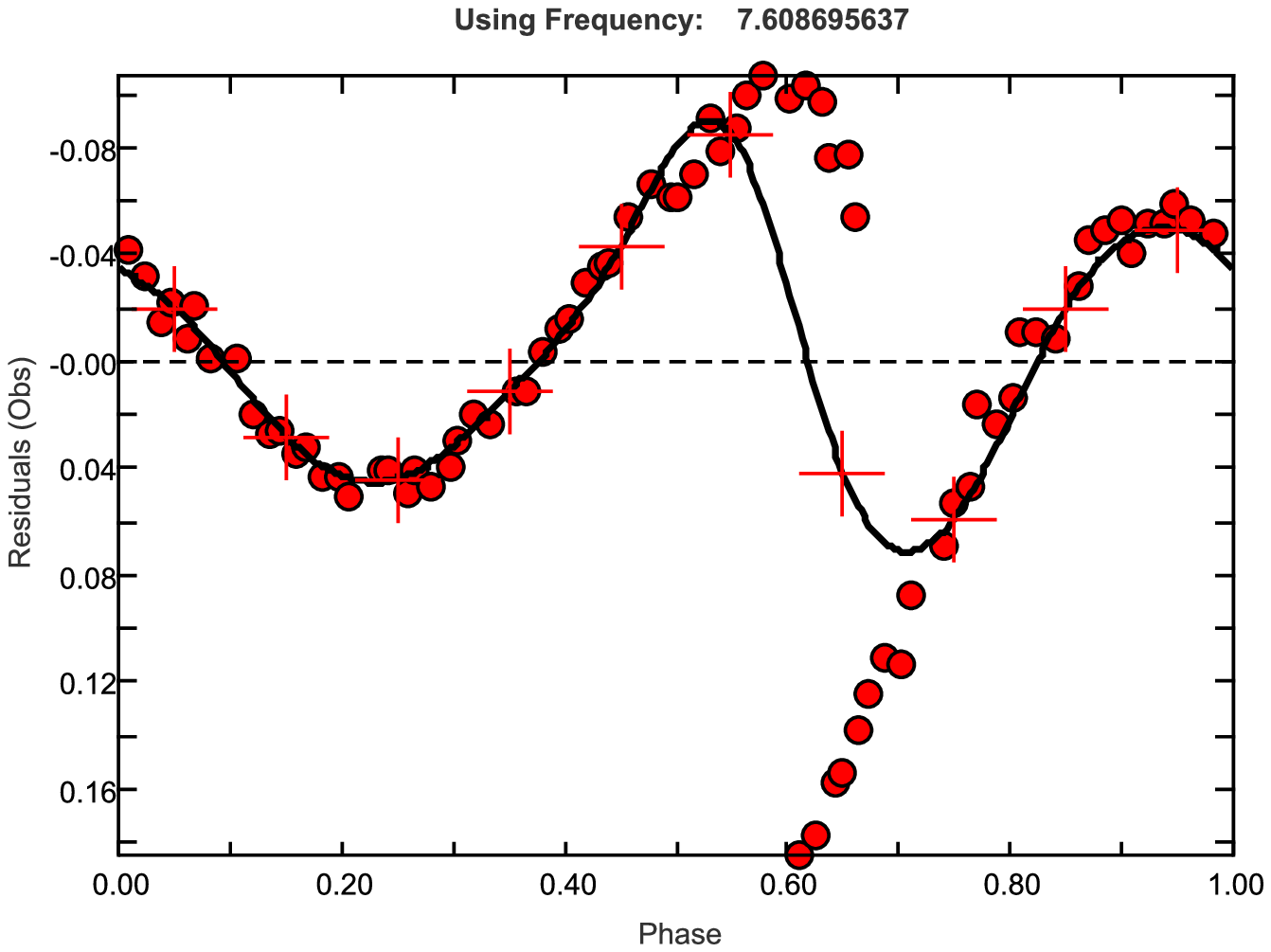}\\
\textbf{{\scriptsize Left: Residual for May 5, 2011 - Right: Residual for May 27, 2011}} \newpage
\includegraphics[width=0.48\textwidth{}]{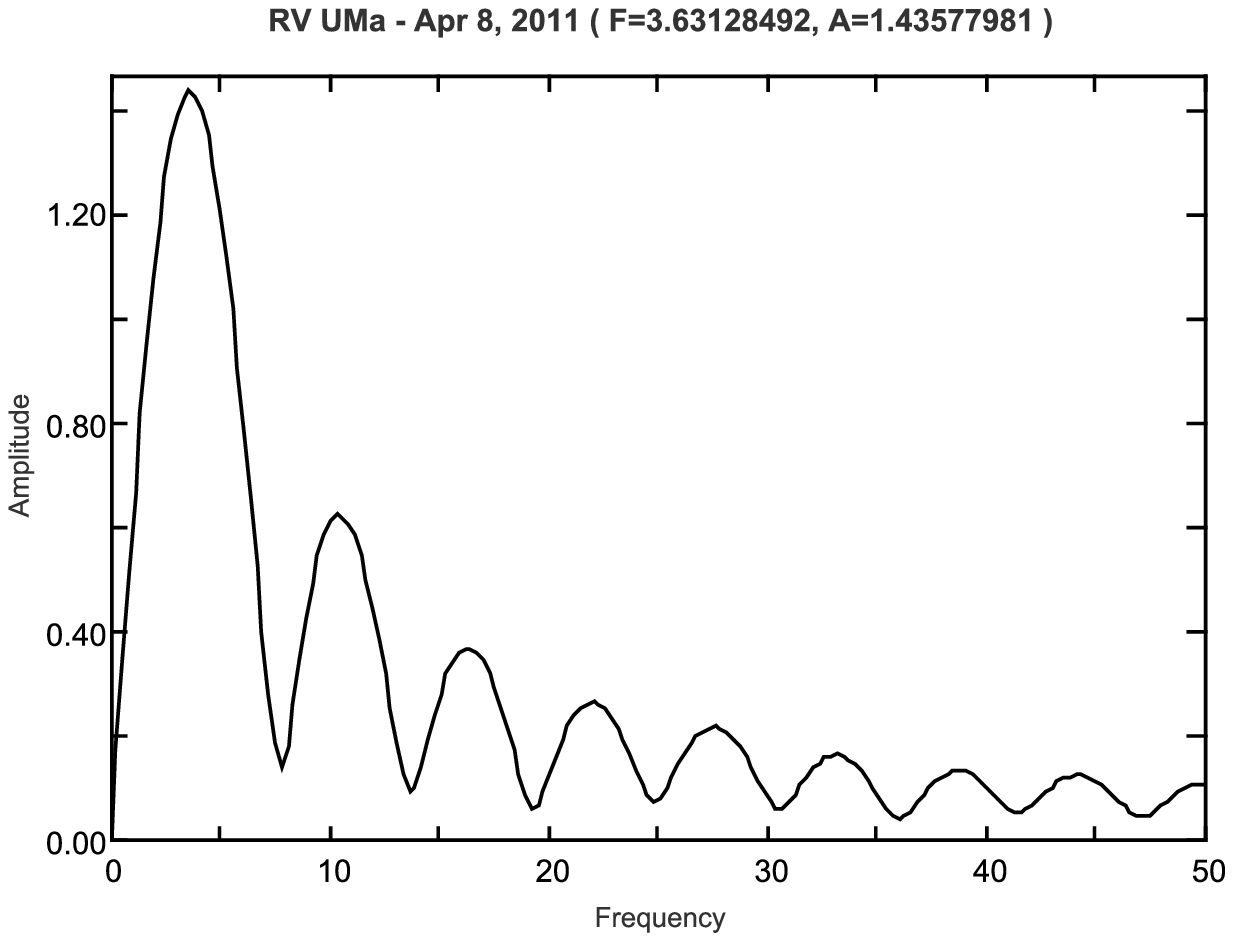}\quad
\includegraphics[width=0.48\textwidth{}]{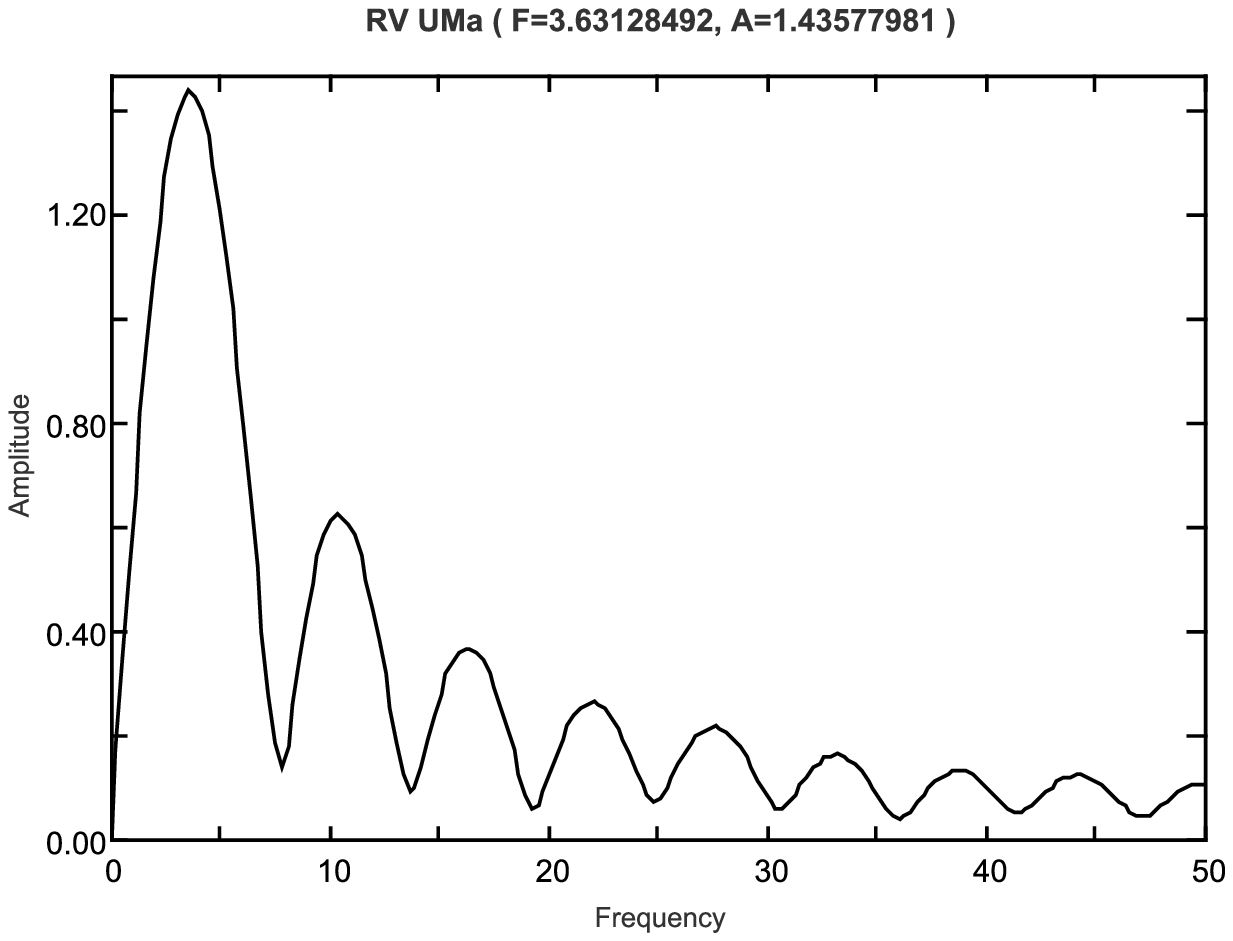}\\ 
\textbf{{\scriptsize Up: Fourier for Apr 8, 2011 - Down: Fourier for Apr 13, 2011}} 
\includegraphics[width=0.48\textwidth{}]{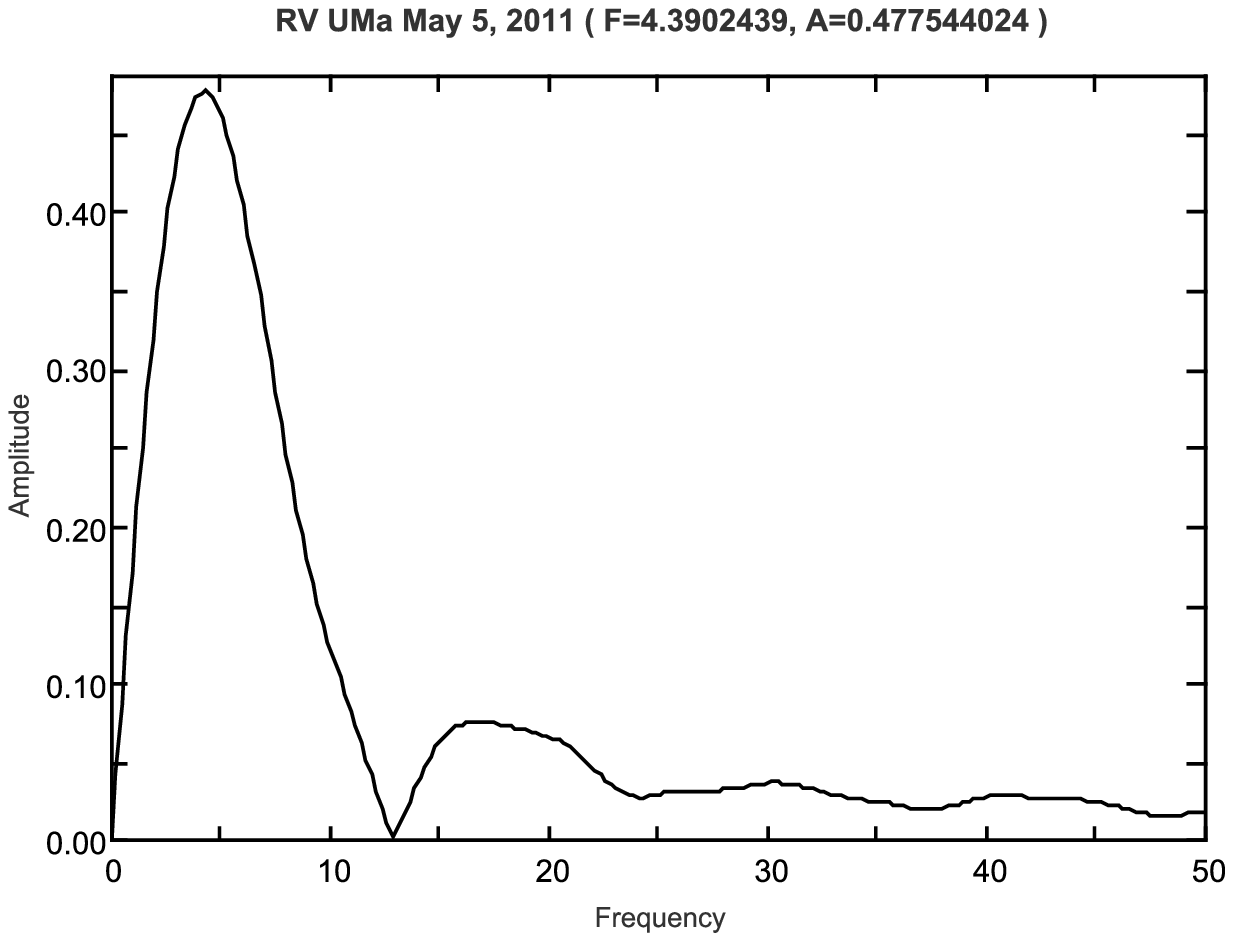}\quad
\includegraphics[width=0.48\textwidth{}]{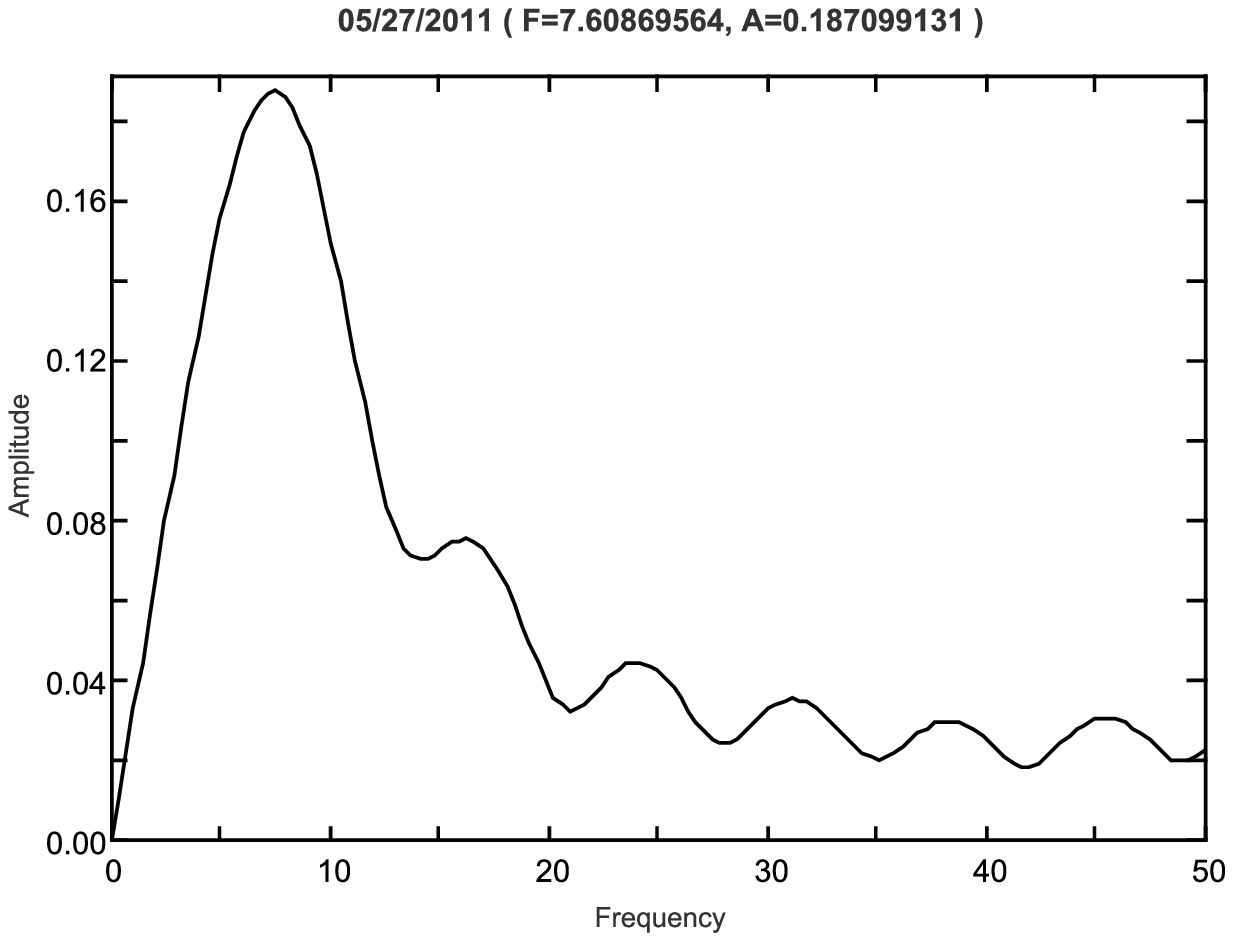}\\
\textbf{{\scriptsize Left: Fourier for May 5, 2011 - Right: Fourier for May 27, 2011}} 

\end{document}